# DEEP LEARNING FOR EXOTIC OPTION VALUATION*

Jay Cao, Jacky Chen, John Hull, Zissis Poulos

**Abstract**

A common approach to valuing exotic options involves choosing a model and then determining its parameters to fit the volatility surface as closely as possible. We refer to this as the model calibration approach (MCA). A disadvantage of MCA is that some information in the volatility surface is lost during the calibration process and the prices of exotic options will not in general be consistent with those of plain vanilla options. We consider an alternative approach where the structure of the user's preferred model is preserved but points on the volatility are features input to a neural network. We refer to this as the volatility feature approach (VFA) model. We conduct experiments showing that VFA can be expected to outperform MCA for the volatility surfaces encountered in practice. Once the upfront computational time has been invested in developing the neural network, the valuation of exotic options using VFA is very fast.

*An earlier version of this paper was titled "Valuing exotic options and estimating model risk"

# DEEP LEARNING FOR EXOTIC OPTION VALUATION

For many underlying assets, there is little uncertainty about the pricing of plain vanilla European and American options. Quotes and trades by market participants provide points on the volatility surface. Interpolating between these points as necessary, a trader can derive a reasonable estimate of the implied volatility appropriate for any new plain vanilla European or American option that is of interest. Plain vanilla options are therefore not priced using a model. They are simply priced to be consistent with the market. The volatility surface derived from the Black–Scholes–Merton model is a convenient interpolation tool for doing this.

Exotic options are generally not as actively traded as plain vanilla options and, as a result, a model is required for pricing. A variety of different models are used in practice. Two conditions that traders would like the model to satisfy are:

A. The stochastic behavior assumed for the underlying asset price should correspond reasonably well to its observed behavior, and
B. The volatility surface derived from the model should be reasonably consistent with the volatility surface used to price plain vanilla options.

Two categories of models that are used in practice can be distinguished. The models in the first category focus on condition A by assuming a process for the asset price that is roughly consistent with its observed behavior. The models have parameters that can be chosen to provide an approximate fit to the current volatility surface. Models in the second category focus on condition B and are designed to be exactly consistent with the current volatility surface.

Many different models in the first category involving stochastic volatility and jumps have been proposed. Examples of stochastic volatility models are Hull and White (1987) and Heston (1993). Merton (1976) proposed a model that overlays Black-Scholes-Merton model with jumps. Bates (1996) adds jumps to Heston (1993). Madan et al (1998) propose a variance–gamma model where there are only jumps. More recently, rough volatility models where the process for volatility is non-Markov have been suggested by authors such as Gatheral et al (2018).

The second category of models are referred to as local volatility models. The original one-factor local volatility model was suggested by Dupire (1994) and Derman and Kani (1994). It has been extended by authors such as Ren et al (2007) and Saporito et al (2019). Local volatility models by design satisfy condition B. However, they are liable to perform less well than models in the first category as far as condition A is concerned.

This paper is concerned with ways in which machine learning can be used to bridge the gap between the two categories of models. It provides a way in which models in the first category can be made more consistent with the volatility surface, and therefore with the pricing of plain vanilla options.

The usual approach to implementing models in the first category is to choose model parameters to fit the volatility surface as closely as possible. This approach, which we refer to as the "model calibration approach" or MCA has been used by researchers such as Horvath et al (2021) and Liu et al (2019). A drawback of the approach is that some of the points on the volatility surface are likely to be more important than others for any particular exotic option that is considered. It is of course possible to vary the weights assigned to different points on the volatility surface according to the instrument being valued. However, it is difficult to determine in advance what these weights should be. As a result, the points are usually given equal weight when model parameters are determined.

Authors such as Ferguson and Green (2018) show how neural networks can be used in conjunction with MCA to provide fast pricing once model parameters have been determined. Consider an exotic option that is valued using Monte Carlo simulation. As a first step, it is necessary to devote computational resources to creating a data set relating model parameters and exotic option parameters to the price. The pricing model can then be replicated with a neural network. Once this has been done, valuation is several orders of magnitude faster than Monte Carlo simulation because it involves nothing more than a forward pass through the neural network.

We suggest an approach, which we refer to as the "volatility feature approach" or VFA. This can be regarded as an extension of the Ferguson and Green (2018) methodology. We create a neural network where the inputs are the volatility surface points and the exotic option parameters and the target is the price. We randomly generate many sets of parameters for (a) the model under consideration and (b) the exotic option under consideration. For each set of parameters, a volatility surface and a price for the exotic option are calculated. The neural network is then constructed. The model parameters are used to create the training set, but they are not inputs to the neural network.

The VFA approach is a compromise between models that satisfy condition A and those that satisfy condition B. It retains some of the structure of an analyst's preferred model while learning which points on the volatility surface are most important for the exotic option being valued. It produces prices for exotic options that are more consistent with the prices of vanilla options than MCA.[1]

We first carry out an experiment to compare the performance of MCA and VFA. We assume that the world is governed by a particular model A, and that the model used for pricing a particular type of exotic option is an alternative simpler model B. The 'true' prices of a panel of exotic options are determined using model A. The average pricing errors of VFA and MCA are about the same in the experiment. However, a key result is that the performance of VFA relative to MCA increases significantly as the MCA calibration error increases. For the volatility surfaces encountered in practice, the calibration error is on average much greater than in our idealized experiment. This provides evidence suggesting that VFA is a better tool than MCA.

As a second test, we compare the pricing of knock-in and knock-out options with vanilla options. A portfolio consisting of a knock-in and a knock-out option with the same barrier level, strike price and maturity should, in the absence of arbitrage, have the same value as a vanilla European option with that strike price and maturity. Using S&P 500 data, we show that this is much closer to being true for VFA than MCA.

Model risk has assumed increasing importance for both dealers and their regulators in recent years. SR 11-7, which was published by the U.S. Board of Governors of the Federal Reserve System in 2011, and is used by other regulators throughout the world, requires banks to identify sources of model risk and carry out frequent model validation exercises.[2] We illustrate how model risk can be assessed using VFA for options on the S&P 500 by comparing Bates (1996) and a version of the rough volatility model.

---

[1] Unlike MCA, the VFA model is not necessarily an arbitrage-free model. This means that there may be theoretical arbitrage opportunities between the pricing of different exotic options. However, this disadvantage is outweighed by the fact that VFA creates less arbitrage opportunities between exotic and vanilla options than MCA.
[2] See Federal Reserve System (2011)

**HESTON AND ITS EXTENSIONS**

In this section we review the Heston (1993) model and two extensions that will be used in our analyses. Define

$S_t$:   Asset price at time $t$

$r$:   Risk-free rate (assumed constant)

$\delta$:   Yield on asset (assumed constant)

The Heston (1993) model is

$$dS_t = (r - \delta)S_t dt + S_t\sqrt{V_t}dz_t$$

$$dV_t = a(V_L - V_t)dt + \xi\sqrt{V_t}\, dw_t$$

The variance rate, $V_t$, reverts to a level $V_L$ at rate $a$. Superimposed on this mean reversion is a 'volatility of volatility' variable, $\xi$. The $dz_t$ and $dw_t$ are Wiener processes with correlation $\rho$.

One extension of the Heston model is Bates (1996). This incorporates jumps in the process for $S_t$. The model is:

$$dS_t = (r - \delta - \lambda\bar{k})S_t dt + S_t\sqrt{V_t}dz_t + kdq$$

$$dV_t = a(V_L - V_t)dt + \xi\sqrt{V_t}\, dw_t$$

The jumps are assumed to occur randomly at rate $\lambda$ per unit time. The percentage jump size, $k$, is assumed to have the property that ln(1+$k$) is normal with mean $\mu$ and standard deviation $\sigma$. The average jump size $\bar{k}$ is $e^{(\mu+\sigma^2)} - 1$.

The rough Heston model, as described by for example El Euch et al (2019), is another extension of the basic Heston model. In this, the process for the variance, $V_t$, involves a fractional Brownian motion rather than a regular Brownian motion. This model can be written as

$$dS_t = (r - \delta)S_t dt + S_t\sqrt{V_t}dz_t$$

$$V_t = V_0 + \frac{1}{\Gamma(H + 0.5)}\int_0^t (t - s)^{H-0.5}\left((a(V_L - V_s)ds + \xi\sqrt{V_s}dw_s\right)$$

where the correlation between $dz_t$ and $dw_t$ is ρ, $H$ is the Hurst exponent, and Γ is the gamma function. When $H$ = 0.5 the model reduces to Heston (1993). When $H$ < 0.5, the correlation between volatility movements in successive periods of time is negative while, when $H$ > 0.5, this correlation is positive. Gatheral et al (2018) produce results for indices showing that the model reflects market data well when $H$ is between 0.06 and 0.20.

A close approximation to the rough Heston model is the lifted Heston model proposed by Jaber (2019). This has the advantage that it involves only regular Brownian motions and, as a result, is much faster for the computation of exotic option prices. We will use the following version of Jaber (2019)

$$dS_t = (r - \delta)S_t dt + S_t\sqrt{V_t}dz_t$$

$$V_t = V_0 + aV_L \sum_{i=1}^{n} \frac{c_i}{x_i}(1 - e^{-x_i t}) + \sum_{i=1}^{n} c_i U_{i,t}$$

$$dU_{i,t} = -x_i U_{i,t} dt + \xi\sqrt{V_t}dw_t$$

where the correlation between the Wiener processes, $dz_t$ and $dw_t$, is ρ. The $U_{i,t}$ are initially zero and revert to zero at different speeds $x_i$.

As recommended by Jaber, we set $n$ = 20 and

$$c_i = \frac{(\beta^{1-\alpha} - 1)\beta^{(\alpha-1)(1+\frac{n}{2})}}{\Gamma(\alpha)\Gamma(2-\alpha)}\beta^{(1-\alpha)i}$$

$$x_i = \left(\frac{1-\alpha}{2-\alpha}\right)\left(\frac{\beta^{2-\alpha} - 1}{\beta^{1-\alpha} - 1}\right)\beta^{i-1-\frac{n}{2}}$$

with α = $H$ + 0.5 and $\beta$ = 2.5.

**A CONTROLLED EXPERIMENT**

In this section, we describe a controlled experiment to compare the results from MCA and VFA. We assume that an asset price evolves according to a particular model A and that a simpler model B is used for pricing a particular type of exotic option dependent on the asset price. The 'true' prices of a panel of exotic options are determined using model A. The prices using both MCA and VFA in conjunction with model B are then computed and the accuracies of the models compared.

We define the volatility surface by considering five different times to maturity (one month, three months, six months, one year, and two years) and five different strike prices (0.7S, 0.85S, S, 1.15S and 1.3S, where S is the asset price). When combined, these maturity/moneyness combinations give 25 'standard' points on the volatility surface. The MCA calibration error is calculated as the minimum value of

$$X = \frac{1}{25}\sum_{j=1}^{25} |\sigma_j - \sigma_j^*| \quad (1)$$

where $\sigma_j^*$ is the 'true' implied volatility calculated from model A and $\sigma_j$ is implied volatility at the $j$th point on the volatility surface given by a set of model B parameters.

The steps in carrying out the controlled experiment are as follows:

(a) Sample $n$ sets of parameters for model A and the exotic option under consideration.
(b) For each of the $n$ scenarios in (a), calculate the 25 implied volatilities that define the 'true' volatility surface.
(c) For each of the $n$ scenarios, calculate the parameters of model B that fit the 25 volatilities calculated in (b) as closely as possible by using an iterative procedure that minimizes the calibration error in equation (1).
(d) Calculate prices for the $n$ exotic options in (a) using both model A and the calibrated model B in (c).
(e) For each of the $n$ scenarios, calculate the MCA error as the absolute value of the difference between the two model prices.
(f) Sample $N$ sets of parameters for model B and the exotic option under consideration. For each, calculate the value of the exotic option and the 25 standard points on the volatility surface.
(g) Use the data from step (f) to construct a neural network where the features are the volatility surface points and the exotic option parameters and the target is the model B exotic option price.
(h) Use the neural network in (g) to calculate the prices of the $n$ exotic options in (a) using the 'true' volatility surface in (b).
(i) For each of the $n$ options in (a), calculate the VFA error as the absolute value of the difference between the price in (h) and the model A price.

The neural networks we used in the experiment consisted of three hidden layers with 30 neurons per layer and used the sigmoid activation function. We used Adam as the optimizer and mean absolute error to measure training loss. In the training process, we saved the weights of the neural network every time when the validation loss reached a new low. We stopped the training when there was no improvement in the validation loss for 50 epochs and used the last saved set of weights as the final weights of the model.

We assumed that model B (the valuation model) is Heston (1993) and model A (the 'true model') is Bates (1996). We considered three different types of exotic options. The first is a knock-out barrier call option where the barrier level is greater than either the initial asset price or the strike price. The second is an Asian option where the payoff is the excess of the average asset price during the life of the option over the strike price if this is positive and zero otherwise. The third is a lookback option where the payoff is the excess of the highest price during the life of the option over the strike price if this is positive and zero otherwise.

We set $n$=1,000 and $N$= 20,000, and used Monte Carlo simulation with 50,000 paths to calculate the prices of the exotic options in (d) and (f).

We assumed an initial asset price, $S_0$, of 100, a yield on the asset, δ, of zero, and sampled other parameters from uniform distributions. We will denote the ranges considered by parentheses. For example, [1, 2] indicates a uniform distribution between 1 and 2.

The model B (Heston) parameters in (f) were sampled as follows

Risk-free rate, $r$: [0.01, 0.05]
Initial variance, $V_0$: [0.01, 0.25]
Mean reversion parameter, $a$: [0.1, 3]
Long-term variance, $V_L$: [0.01, 0.25]
Volatility of volatility, $\xi$: [0.1, 0.8]
Correlation, ρ: [−0.9, 0]

The model A (Bates) parameters in (a) were sampled similarly with the following additional samples:
Jump intensity, λ: [1, 5]
Jump size parameter, $\mu$: [−0.05, 0.05]
Jump size parameter, σ: [0, 0.05]

The option parameters in (a) and (f) were sampled as follows:

Time to maturity (yrs): [0.05, 2]
Strike price: [80, 120]

For barrier options, the barrier level was set equal to $C$ times the maximum of the asset price and strike price with $C$ sampled from [1.05, 1.30].

Define $Y$ as the MCA minus the VFA error. We plotted $Y$ against the calibration error, $X$, in equation (1). The results are shown in Exhibit 1. Overall VFA performed only slightly better than MCA, but the $t$-statistics indicate that as the calibration error increases the performance of VFA relative to MCA improves significantly. This result is not too surprising. When the model provides a good fit to the volatility surface MCA performs well. As the fit becomes worse, VFA starts to perform relatively better because it retains all information about the volatility surface.

**Exhibit 1**: Results of regressing the MCA error minus the VFA error (Y) against the MCA calibration error (X). $t$-statistics are in parentheses.

| Exotic Option | Relationship |
|---|---|
| Barrier | $Y = -0.10 + 16.35X$ <br> (−8.30)   (15.64) |
| Asian | $Y = -0.21 + 10.37X$ <br> (−17.90)   (9.96) |
| Lookback | $Y = -0.42 + 58.42X$ <br> (−11.41)  (18.04) |

The results are encouraging. The Bates model is an extension of Heston and shares many of its properties. The volatility surfaces encountered in practice can be expected to be 'less well behaved' than those derived from the Bates model. To investigate this, we fitted Heston to S&P 500 volatility surfaces between 2001 and 2019. As will be described later, the volatility surface was defined using 19 rather than 25 standard points, and the points were determined from market data using a bivariate linear interpolation.

We found the S&P 500 calibration errors to be over twice as high on average as the Bates calibration errors. Given that the relative performance of VFA improves as the calibration error increases, our results suggest that VFA is likely to be a better tool than MCA in practice.[3] Once the upfront computational time to create the neural network has been invested, VFA provides a fast pricing tool.[4]

**IMPORTANCE OF VOLATILITY SURFACE POINTS**

We implement MCA by minimizing

$$\frac{1}{25}\sum_{j=1}^{25}|\sigma_j - \sigma_j^*|$$

where $\sigma_j$ is the $j$th point on the volatility surface given by model B and $\sigma_j^*$ is the 'true' volatility at the $j$th on the volatility surface. This objective function (in common with other similar objective functions that might be used such as mean squared error) assumes that the 25 points are equally important. In this section we outline analyses we have carried out showing that the points are not equally important for any particular option or when averaged across all the options that are considered. This suggests that, for the valuation of any given option, some relevant information is lost when MCA is used and may be a reason why VFA tends to give more accurate results as the MCA calibration error increases.

The sensitivity of the value of exotic options to the positions of the 25 standard points on the volatility surface can be estimated from the gradient of the function that the VFA neural network represents. Letting $y$ represent the exotic option price estimated by the neural network, we define the sensitivity, $s(v_i)$, of the price to the implied volatility at the $i$th point on the volatility surface as the absolute value of

---

[3] The fact that 25 points had to be fitted in the experiments and only 19 points had to be fitted to the S&P 500 data reinforces our conclusions.
[4] As shown by Horvath et al (2021), neural networks can be used to speed up MCA computations.

$$\frac{\left|\frac{\partial y}{\partial v_i}\right|}{E\left(\left|\frac{\partial y}{\partial v_i}\right|\right)} - 1$$

where $\frac{\partial y}{\partial v_i}$ is the partial derivative of the prediction function with respect to volatility surface point $v_i$.

In our sensitivity analysis we aim to measure the rate of change irrespective of sign, thus we only consider the absolute values of partial derivatives. Further, in the equation above, expectation is taken over the entire set of volatility surfaces that are used. Computing the average partial derivative with respect to each volatility surface point is necessary so that the baseline sensitivity of the neural network is taken into account and used as reference. A detailed discussion on the use of baselines for gradient-based sensitivity analysis can be found in the work of Sundararajan et al (2017). Intuitively, the average gradient corresponds to how sensitive the neural network output is when the 'average' surface is passed as input. We therefore define sensitivity to $v_i$ using the absolute relative difference between $\left|\frac{\partial y}{\partial v_i}\right|$ and its average: a larger sensitivity implies a larger impact on how the predicted value changes relative to the expected output of the network.

The gradients are computed by applying the back-propagation algorithm, which was first introduced by Rumelhart et al (1986) and is typically used to train neural networks. Specifically, we apply back-propagation on the final trained network and take an additional step in the chain rule to obtain input gradients. Note that during this process we do not update the neural network weights.

We randomly sampled 200 volatility surfaces and calculated $s(v_i)$ for barrier, Asian, and lookback options for $i$=1 to 25. Exhibit 2 reports the standard deviations of the $s(v_i)$. From this exhibit, it can be seen that there is variability in how a particular point contributes to the price prediction across different input surfaces. These and other more detailed results allow us to conclude that a) different volatility surface points have different importance when valuing a single exotic option, and b) the same volatility point has varying importance when valuing different exotic options. The results corroborate our hypothesis that the VFA approach, by virtue of using a neural network, can exploit the structural information of the volatility surface when valuing an exotic option and can therefore be advantageous compared to MCA.

**Exhibit 2:** Standard deviation of sensitivity of price to implied volatility for barrier, Asian and lookback options. *T* is the maturity for volatility surface point considered, *K* is strike price and *S* is asset price.

|  | Barrier Options | | | | |
| --- | --- | --- | --- | --- | --- |
|  | T = 1 month | T = 3 months | T = 6 months | T = 1 year | T = 2 years |
| K = 0.7S | 0.77 | 0.59 | 0.22 | 0.47 | 0.37 |
| K = 0.85S | 0.37 | 0.52 | 0.42 | 0.37 | 0.48 |
| K = S | 0.58 | 0.45 | 0.33 | 0.34 | 0.47 |
| K = 1.15S | 0.49 | 0.40 | 0.38 | 0.40 | 0.32 |
| K = 1.3S | 0.36 | 0.42 | 0.38 | 0.46 | 0.25 |
|  | Asian Options | | | | |
| K = 0.7S | 0.67 | 0.65 | 0.40 | 0.39 | 0.40 |
| K = 0.85S | 0.26 | 0.44 | 0.34 | 0.43 | 0.54 |
| K = S | 0.37 | 0.23 | 0.35 | 0.38 | 0.47 |
| K = 1.15S | 0.56 | 0.44 | 0.31 | 0.38 | 0.34 |
| K = 1.3S | 0.30 | 0.35 | 0.27 | 0.48 | 0.31 |
|  | Lookback Options | | | | |
| K = 0.7S | 0.40 | 0.22 | 0.34 | 0.44 | 0.57 |
| K = 0.85S | 0.40 | 0.22 | 0.34 | 0.31 | 0.41 |
| K = S | 0.51 | 0.23 | 0.29 | 0.28 | 0.41 |
| K = 1.15S | 0.58 | 0.23 | 0.24 | 0.29 | 0.40 |
| K = 1.3S | 0.33 | 0.26 | 0.27 | 0.43 | 0.52 |

It is clear from the analysis in this section that it is at best an approximation to give all points on the volatility surface equal weights when an exotic option is valued. The weights appropriate for different points vary according to the parameters of the exotic being considered. When MCA is used the volatility surface is fitted with some error. It is possible to vary the weights used for different points on the volatility surface so that some parts of the volatility surface are fitted more accurately than other parts. However, it is difficult to know ex ante what the appropriate weights are. This is what motivates us to experiment with the VFA approach. The neural network that is used in VFA has the potential to better reflect inter-relationships between points on the volatility surface and the parameters of the exotic option being valued.

**S&P 500 HISTORICAL DATA**

The rest of our tests involve S&P 500 data. We collected data on call options on the S&P 500 between June 2001 and June 2019 from OptionMetrics. We cleaned the data in a number of ways. In particular, we only kept options with open interest greater than 0 and time to maturity between 1 month and 2 years. Options with no implied volatility reported by the database were removed. We defined moneyness as the strike price divided by the index level and removed all options with moneyness smaller than 0.7 and greater than 1.3. This led to an average number of options each day of 500.4.

In defining the S&P 500 volatility surface, only 19 of the 25 points used in the controlled experiments were used. The one-month implied volatilities for moneyness levels of 0.7, 0.85, 1.15, and 1.3, and the three-month implied volatilities for moneyness levels of 0.7 and 1.3 were not used. This is because S&P 500 implied volatilities were often either unreliable or nonexistent for these extreme moneyness/maturity combinations.

Options trade on any given day with nonstandard maturity/moneyness combinations. Each day we used a search algorithm to determine the implied volatilities for standard points that, with interpolation, gave the best fit to the options in our data set. Consider a particular option in the data set has a strike price of $K$ and time to maturity $T$. Define $K_u$ and $K_d$ as the standard strike prices that are closest to $K$ with the property that $K_u \geq K \geq K_d$. Similarly, define $T_u$ and $T_d$ as the standard times to maturity that are closest to $T$ with the property that $T_u \geq T \geq T_d$. For any given trial set of standard implied volatilities, the implied volatility for the $\{K, T\}$ option was determined using a bivariate linear interpolation between the $\{K_u, T_u\}$, $\{K_u, T_d\}$, $\{K_d, T_u\}$, and $\{K_d, T_d\}$ implied volatilities. The best-fit implied volatilities for the standard maturities and strike prices were those that minimized the sum of squared differences between the interpolated volatilities and the reported implied volatilities.

**CHOOSING MODEL PARAMETERS**

For computational efficiency, it is important that the model parameters sampled to generate the training set lead to volatility surfaces that have similar characteristics to those that have been observed for the underlying asset. Mutual information measures can be used to test this.[5]

The first step is to create a balanced data set where half the volatility surfaces come from historical data and half come from sampled model parameters. The k-means algorithm is then used to partition the data into two clusters. Finally, adjusted mutual information (AMI) is used to measure the quality of the produced clusters. A large AMI indicates good clustering, where sampled and historical surfaces are mostly placed in separate clusters, i.e., it is easier for the algorithm to discriminate between the two types of surfaces. The opposite is also true. If AMI is small then it is harder to discriminate between the two types of surfaces, and thus one can conclude that they share more similar characteristics. Based on this rationale we can determine whether the sampled surfaces were similar enough to the historical surfaces using the following rules:

(a) If the average AMI of clustering is distinctly larger than the average AMI of a random partition, then the sampled surfaces are unsatisfactory for capturing the data distribution of S&P volatility surfaces. In other words, sampled surfaces were easy to identify and the model parameters should be adjusted.
(b) If the average AMI of clustering was close to the average AMI of a random partition, then the sampled surfaces were considered valid for our purposes and the model parameters were accepted.

Note that this process was not used to explicitly search for the best model parameters (clustering results did not dictate how model parameters should be sampled), but rather it was applied as a validation step to ensure the quality of the dataset

In the case of the S&P 500, the at-the-money one-month implied volatility, $X$, ranged from 0.054 to 0.7589. The value of $\ln(X-0.05)$ had a mean of $-2.41$ and a standard deviation of 0.69. This led us to sample the initial volatility for the Heston model and its extensions as $0.05+\exp(u)$ where $u$ has a mean of −2.5 and a standard deviation of 1.0. The best fit relationship between the two-year implied volatility, $W$, and the one-month implied volatility, $Z$, for the S&P 500 was

$$W = 0.12 + 0.46Z$$

with a standard error of 0.02. This led us to sample the long-term variance rate so that the two-year volatility was in a 0.08 range that was approximately centered on this. Other parameters were sampled similarly to the controlled experiments. This led to the sampled volatility surfaces satisfying our AMI test.

---

[5] See Vinh et al. (2010) for a discussion of mutual information measures.

**BARRIER OPTION TEST**

Define $V_{KO}$ and $V_{KI}$ as the value of a knock-out and knock-in options with barrier $H$, strike price $K$, and time to maturity $T$. For no arbitrage we must have:

$$V_{KO} + V_{KI} = V$$

where $V$ is the price of a vanilla European option with strike price $K$ and time to maturity $T$. This relationship enables us to provide a test of the extent to which exotic option pricing is consistent with vanilla option pricing.

For each day for which we had S&P 500 data we created knock-in, knock-out and vanilla options with a common maturity, strike price, and barrier. The maturities were chosen by randomly sampling from four alternatives: three months, six months, one year and two years. The strike prices were then chosen by randomly sampling from a uniform distribution as indicated in Exhibit 3. For example, the one-year options have strike prices between 85% and 115% of the index. Finally, a barrier multiplier, $\beta$, was sampled from a uniform distribution as indicated in Exhibit 3. The barrier was then set equal to $\beta$ times $\max(S, K)$ where $S$ is the index level and $K$ is the strike price. For example, one-year options had barriers that were between 120% and 135% of $\max(S, K)$.

**Exhibit 3:** Strike price and barrier ranges

| Maturity | Strike Price (% of S&P 500) | Barrier Multiplier, $\beta$ |
|---|---|---|
| 3 months | [92.5,107.5] | [1.10,1.15] |
| 6 months | [90,110] | [1.10,1.30] |
| 1 year | [85,115] | [1.20,1.35] |
| 2 years | [80,120] | [1.30,1.50] |

The volatility surfaces each day were calculated using the double-linear interpolation method described earlier. These volatility surfaces were then used with interpolation to calculate an implied volatility, and therefore a market price for the vanilla options.

To value the barrier options we used MCA and VFA in conjunction with Heston (1993). For MCA, we determined Heston parameters by minimizing the mean absolute error between the calibrated and the historical volatilities.[6] We then used Monte Carlo simulation to generate asset paths and calculate the prices of the knock-in and knock-out options.

For VFA, we sampled 400,000 sets of model parameters and barrier option parameters. The model parameters were sampled as explained in the previous section. A neural network was then used to relate knock-in and knock-out option prices to (a) volatility surface points and (b) the option parameters. The neural networks were then used to calculate the prices of the options than had been sampled each day.

MCA and VFA prices are compared in Exhibit 4. On average, the VFA prices of knock-in and knock-out barrier options are more consistent with the vanilla price by an amount which is 0.44% of the option price. The MCA pricing consistency deteriorates as the maturity of the option increases while the VFA performance remains about the same.

**Exhibit 4:** Mean absolute error of $V_{\text{KO}} + V_{\text{KI}} - V$. All values are measured as a percent of the S&P 500 index.

| T | MCA MAE | VFA MAE | Sample Size |
|---|---|---|---|
| 0.25 | 0.24 | 0.12 | 1110 |
| 0.5 | 0.24 | 0.11 | 1059 |
| 1 | 0.50 | 0.09 | 1065 |
| 2 | 1.21 | 0.14 | 1066 |
| Full Sample | 0.55 | 0.11 | 4300 |

**MODEL RISK**

Model risk can be quantified using VFA. We illustrate this by comparing the prices given by two models for exotic options on the S&P 500. The first model is Bates (1993). The second is the approximation to rough Heston proposed by Jaber (2019).

---

[6] We eliminated data for 351 of the 4,651 sample days because our calibration algorithm failed on those days. Our results therefore conservative in that they reflect only days for which MCA is a plausible approach.

The steps used to produce results for each of the models and exotic options considered are as follows:

(a) Model parameters are generated randomly as discussed in the previous section. For each set of model parameters, a set of parameters describing the exotic option are also generated randomly. For each of the resulting data sets, the standard points on the volatility surface are calculated and the exotic option is valued.

(b) Data from step (a) are used to construct a neural network. The features are the volatility surface points and the exotic option parameters while the target is the exotic option price.

(c) The neural network in (b) is used to price a panel of exotic options using historical data on the asset of interest (the S&P 500 in our case)

The price differences when different models are used to price the same panel of exotic options provide statistics to quantify model risk

Exhibit 5 shows the means and standard deviations of the difference between the prices given by rough Heston and Bates for the three exotic options considered. It can be seen that rough Heston tends to produce higher prices than Bates for barrier and Asian options and lower prices for lookback options.[7]

**Exhibit 5:** Statistics for the difference between rough Heston (RH) and Bates prices, as a percentage of the index level

|  | Mean RH Price | Mean Bates Price | Mean Price Diff: RH minus Bates | S.D. of Price Difference |
|---|---|---|---|---|
| Barrier | 4.494 | 4.286 | 0.208 | 0.271 |
| Asian | 4.880 | 4.835 | 0.045 | 0.154 |
| Lookback | 11.631 | 11.946 | −0.315 | 0.484 |

We also tested whether pricing differences depend on the calibration error when Heston is fitted to the S&P 500 volatility surfaces. For all three exotic options we found that price differences increase significantly as the calibration error increases. This suggests (unsurprisingly) that model risk increases as the deviation of the volatility surface from the volatility surfaces given by models commonly used increases.

---

[7] A reason for this may be that the jumps in Bates make lookbacks more valuable and the barrier more likely to be hit.

**CONCLUSIONS**

In this paper, we have contrasted two approaches for valuing exotic options. One widely used approach, MCA, involves fitting a model to the current volatility surface as closely as possible. The other approach, VFA, uses the model in a different way. It relates points on the volatility surface to the price and allows a neural network to learn which points on the volatility surface are most relevant for valuing a particular exotic option. Tests show that VFA provides an interesting alternative to local volatility models for analysts who are concerned that MCA leads to arbitrage opportunities between exotic options and vanilla options.

# References

Bates D. S. 1996. "Jumps and Stochastic Volatility: Exchange Rate Processes Implicit in Deutschemark Options." *Review of Financial Studies* 9 (1): 69-107.

Derman, E. and I. Kani. 1994. "Riding on a Smile." Risk (February): 32-39.

Dupire B. 1994. "Pricing with a Smile." *Risk*. (January):18-20.

El Euch O., J. Gatheral, and M. Rosenbaum. 2019. "Roughening Heston." *Risk* (May): 84-89.

Federal Reserve System. 2011. "SR Letter 11-7: Supervisory Guidance on Model Risk Management." Washington, DC.

Ferguson R. and A. Green. 2018. "Deeply Learning Derivatives." Working paper, arXiv: 1809.02233.

Gatheral J., T. Jaisson, M. Rosenbaum. 2018. "Volatility is Rough." *Quantitative Finance.* 18 (6):933-949.

Heston S. L. 1993. "A Closed Form Solution for Options with Stochastic Volatility with Applications to Bond and Currency Options." *Review of Financial Studies.* 6 (2): 327-343.

Horvath B., A. Muguruza, M. Tomas. 2021. "Deep Learning Volatility: A Deep Neural Network Perspective on Pricing and Calibration in (Rough) Volatility Models." *Quantitative Finance.* 21 (1): 11-27.

Hull, J. C. and A. White. 1997. "The Pricing of Options on Assets with Stochastic Volatility." *Journal of Finance.* 42 (2): 281-300.

Jaber E. A. 2019. "Lifting the Heston model." *Quantitative Finance.* 19 (12): 1995-2013.

Liu S., A. Borovykh, L. A. Grzelak, C. W. Oosterlee. 2019. "A Neural Network-based Framework for Financial Model Calibration. *Journal of Mathematics in Industry.*
9(9), https://doi.org/10.1186/s13362-019-0066-7.

Madan, D. B., P. P. Carr, and E. C. Chang. 1988. "The Variance-Gamma Process and Option Pricing." *European Finance Review*. 2: 79-105.

Merton R. C. 1976. "Option Pricing when Underlying Stock Returns are Discontinuous." *Journal of Financial Economics.* 3 (1-2): 125-144.

Ren, Y., D. Madan, and M. Q. Qian (2007), "Calibrating and Pricing with Enbedded Local Volatility Models." *Risk.* (September): 138-143.

Saporito, Y. F., X. Yang, and J. P. Zubelli. 2019. "The Calibration of Stochastic Local Volatility models: An Inverse Problem Perspective." *Computers and Mathematics with Applications.* 77 (12): 3054-3067.

Sundararajan M., A. Taly, and Q. Yan. 2017. "Axiomatic Attributions for Deep Networks." *Proceedings of 34th International Conference on Machine Learning*: 3319-3328.

Vinh, N. X., J. Epps, and J. Bailey. 2010. "Information Theoretic Measures for Clusterings Comparison: Variants, Properties, Normalization and Correction for Chance." *Journal of Machine Learning Research*. 11: 2837–2854.